\documentclass[aps,pra,twocolumn,showpacs,amsmath,floatfix,a4paper]{revtex4}

\usepackage{graphicx}
\usepackage{bbm}
\usepackage{latexsym}

\newcommand{\identity}{\mathbbm{1}}
\newcommand{\bra}[2][]{\mathinner{\raisebox{-0.3em}
{\ensuremath{\scriptstyle #1}}\hspace{-0.1em}\langle #2|}}
\newcommand{\ket}[2][]{\mathinner{|#2\rangle}_{\hspace{-0.1em}#1}}
\newcommand{\Bra}[2][]{\raisebox{-0.3em}
{\ensuremath{\scriptstyle{#1}}}\hspace{-0.2em}\left<#2\right|}
\newcommand{\Ket}[2][]{\left|#2\right>_{\hspace{-0.1em}#1}}
\newcommand{\braket}[2]{\mathinner{\langle #1|#2\rangle}}
\newcommand{\ketbra}[3][]{\mathinner{|#2\rangle
\raisebox{-0.3em}{\ensuremath{\scriptstyle #1}}\langle #3|}}
\newcommand{\Braket}[2]{\left< #1|#2\right>}
\newcommand{\Ketbra}[3][]{\left|#2\vphantom{#3}\right>\hspace{-0.2em}
\raisebox{-0.3em}{\ensuremath{\scriptstyle #1}}\hspace{-0.2em}
\left<#3\vphantom #2\right|}
\newcommand{\proj}[2][]{\ketbra[#1]{#2}{#2}}
 
\newcommand{\abs}[1]{|#1|} \newcommand{\Abs}[1]{\left\vert #1
\right\vert} \newcommand{\matnorm}[1]{||#1||}
\newcommand{\Matnorm}[1]{\left\Vert #1\right\Vert}

\DeclareMathOperator{\tr}{tr}

\DeclareMathOperator{\real}{Re}
\DeclareMathOperator{\imag}{Im}
\newcommand{\dd}{\mathrm{d}}
\newcommand{\order}[1]{O({#1})}
\newcommand{\Order}[1]{O\left({#1}\right)}
\def\mathclap{\mathpalette\mathclapinternal}
\def\mathclapinternal#1#2{\clap{$\mathsurround=0pt#1{#2}$}}

\def\clap#1{\hbox to 0pt{\hss#1\hss}}

\def\mathrlap{\mathpalette\mathrlapinternal}
\def\mathclap{\mathpalette\mathclapinternal}

\def\mathrlapinternal#1#2{\rlap{$\mathsurround=0pt#1{#2}$}}
\def\mathclapinternal#1#2{\clap{$\mathsurround=0pt#1{#2}$}}

\newcommand{\dt}{\delta t}
\newcommand{\dU}{U_{\dt}}

\newcommand{\hc}{\mathrm{h.c.}}
\newcommand{\bipart}[2]{(#1:#2)}
\newcommand{\etal}{\textit{et.\ al}}

\newtheorem{thm}{Theorem}
\newenvironment{prf}{\noindent \textbf{Proof}}{$\Box$}
\newtheorem{lem}[thm]{Lemma}
\newtheorem{prop}[thm]{Proposition}

\begin{document}

\title{Entanglement flow in multipartite systems}
\date{April 30, 2004}
\author{T. S. Cubitt}\email{toby.cubitt@mpq.mpg.de}
\author{F. Verstraete}\email{frank.verstraete@mpq.mpg.de}
\author{J.I. Cirac}\email{ignacio.cirac@mpq.mpg.de}
\affiliation{Max Planck Institut f\"ur Quantenoptik, Hans--Kopfermann
Str.\ 1, D-85748 Garching, Germany}


\begin{abstract}
  We investigate entanglement dynamics in multipartite systems,
  establishing a quantitative concept of \emph{entanglement flow}:
  both flow through individual particles, and flow along general
  networks of interacting particles. In the former case, the rate at
  which a particle can transmit entanglement is shown to depend on
  that particle's entanglement with the rest of the system. In the
  latter, we derive a set of \emph{entanglement rate equations},
  relating the rate of entanglement generation between two subsets of
  particles to the entanglement already present further back along the
  network. We use the rate equations to derive a lower bound on
  entanglement generation in qubit chains, and compare this to
  existing entanglement creation protocols.
\end{abstract}

\maketitle

\section{Introduction}
New fields of physics often give rise to new physical quantities to
study, and quantum information theory has proved a rich source of
study material. As an amalgam of quantum mechanics and information
theory, many of the new quantities are quantum analogues of familiar
friends from classical information theory: the \emph{qubit}, for
instance, measures quantum information just as the \emph{bit} measures
classical information~\cite{Sch95}. Other quantities have no obvious
classical counterpart. The best-known example is entanglement.
Originally seen as the phenomenon that epitomized quantum weirdness,
it has become established over the last decade as as a physical
quantity, on a par with, say, energy.

The analogy with energy can be pushed quite far: entanglement has a
number of similar properties. Like energy, entanglement can be
quantified in a meaningful way~\cite{QIC}, allowing us to say that one
state is more entangled than another. Like energy, entanglement can be
converted from one form to another~\cite{Nie99}. And like energy, it
is a resource that can be used to carry out useful tasks, such as
teleportation~\cite{BBC+95}.

Until recently, work concentrated on understanding these static
properties of entangled quantum states. Although we are some way from
a complete understanding of entanglement statics, there has been
significant progress: for instance, bipartite pure-state entanglement
is now well understood. This begs the question: what happens if we
allow the state to evolve?

The move from entanglement statics to entanglement \emph{dynamics}
raises many new and interesting questions. How does entanglement
evolve as particles interact~\cite{DVC+01}? How good is a particular
interaction at creating entanglement~\cite{DVC+01,BHWS03,CLVV03}? More
generally, how good is an interaction at simulating various non-local
processes~\cite{KBG01,BDNB04}? Or, turning this on its head, how
`non-local' is a given process (e.g.\ a quantum
gate)~\cite{VHC02,ZZF00}? This article extends the first of these ---
how entanglement evolves as particles interact --- to multipartite
systems.

The Schr\"odinger equation already implicitly describes the complete
dynamics of a quantum system, but to gain insight into entanglement
dynamics, we need equations that \emph{explicitly} involve the
entanglement of the system, without reference specific features of the
Hamiltonian. One of the first steps along this path was taken by D\"ur
\etal.\, who investigated the rate of entanglement generation in
two-qubit systems~\cite{DVC+01}. They derived an equation relating the
rate of entanglement creation to the existing entanglement in the
system, along with a factor depending on the form and strength of the
interaction. This latter led to a pleasingly simple quantity measuring
the entanglement generating capacity of two-qubit
interactions~\cite{BHWS03,CLVV03}.

In a system of two particles coupled by a Hamiltonian, the only
entanglement dynamics that can take place is creation of entanglement
between the two particles. A simple tripartite system already raises
other interesting questions. For instance, in a chain of three
particles, how does entanglement `flow' through the middle one?
Surprisingly, we showed in previous work~\cite{CVDC03} that, in just
such a chain, entanglement can be created between the two end
particles, without the middle particle \emph{ever} becoming
entangled. This would seem to put an end to notions of entanglement
`flow'. However, we also gave a simple proof that this phenomenon is
only possible for mixed initial states; for pure states, the middle
particle necessarily becomes entangled during the evolution.

This suggests there is a connection between pure-state entanglement of
a mediating particle and entanglement flow through that particle: if
it is not entangled, no entanglement flows. In
section~\ref{sec:through}, we show that there is indeed a quantitative
relation describing how the entanglement of a particle with the rest
of the system limits the flow of entanglement through that particle.
We first consider a three-qubit system, before dealing with general
systems. The concept of entanglement flow \emph{through} particles is
therefore put on a quantitative footing for systems in pure states.
This contrasts strongly with the mixed-state case, in which
entanglement can seemingly `tunnel' through mediating systems.

Flow through individual particles is one aspect of entanglement
dynamics in multipartite systems. But in a network of many interacting
particles, we may also be interested in how entanglement flows along
the whole network. We develop these ideas in the second half of this
article.

The inspiration is loosely based on the Arrhenius equations for
chemical reactions. The reaction mechanism of a chemical reaction
describes the steps by which reactants are transformed, via successive
intermediate compounds, into the final products. The rate at which a
compound is produced depends on the amounts of its immediate
precursors that are present. Thus the complete reaction is described
by a set of coupled rate equations, one for each step in the reaction
mechanism.

In section~\ref{sec:along}, we derive a set of differential equations
describing entanglement flow, analogous to the rate equations for a
chemical reaction. The rate at which entanglement is generated between
two sets of particles is shown to depend on the amount of entanglement
already present further back along the network. The entanglement
dynamics of the complete system is described by a coupled set of such
entanglement rate equations, one for each step in the interaction
network.

Unlike the equations describing flow through individual particles,
these entanglement rate equations apply equally well to both pure and
mixed states. Therefore, they establish a concept of entanglement flow
along general networks of interacting particles (even though the
concept of flow \emph{through} individual particles in the system may
be meaningless).

In section~\ref{sec:how_fast}, we apply our new understanding of
entanglement flow to investigate entanglement generation in chains of
interacting particles. First, we briefly review some existing
entanglement generation protocols for qubit chains, in the context of
the rate equations derived in section~\ref{sec:along}. Finally, we use
the rate equations to prove a universal lower bound on the time it
takes to create entanglement, or more precisely the scaling of this
with the length of the chain (the results can easily be extended to
general networks).


\section{Flow through particles\label{sec:through}}
In this section, we will investigate entanglement flow through
mediating particles. Specifically, we will consider flow through the
middle particle in tripartite chains. The results of~\cite{CVDC03}
show that this concept does not make sense if the whole system is in a
mixed state. But for pure states, the rate at which entanglement is
generated between the end particles is indeed zero if the middle
particle is not entangled.

The latter is suggestive: is there a general quantitative relationship
between the entanglement of a particle, and the rate at which
entanglement can flow through it, for systems in pure states? If the
middle particle is only slightly entangled, does entanglement flow
only slowly? We will derive just such a relationship, first for the
simplest tripartite system: a three-qubit chain, then for general
tripartite chains.

When investigating (bipartite) entanglement flow through mediating
particles in more general settings, the system can always be described
as a tripartite chain: the mediating particles form one party, and the
sets of particles each side, which are becoming entangled, form the
other two. Thus the equation for tripartite chains can in fact be
applied generally to describe entanglement flow through mediating
systems.

\subsection{The three-qubit chain}
Consider a chain of three qubits, labeled $a$, $b$, and $c$, with
nearest neighbour interactions described by Hamiltonians $H_{ab}$ and
$H_{bc}$. 
We will restrict the overall state of the system, $\ket[abc]{\psi}$ to
be pure. However, the reduced state of the two end qubits,
$\rho_{ac}$, need not remain pure during the evolution (if $b$ is to
become entangled at any point, $\rho_{ac}$ will necessarily become
mixed). To quantify the entanglement between $a$ and $c$, we need an
entanglement measure valid for mixed states. The natural choice is the
concurrence~\cite{Woo98}. Though it is an entanglement measure in its
own right, its interest lies in its is equivalence to one of the
important, physically meaningful entanglement measures: the
entanglement of formation~\cite{BDSW96}.


We can write the overall state of the system in its Schmidt
decomposition with respect to the partition $\bipart{b}{ac}$:
$\ket[abc]{\psi} = \lambda_1\ket[ac]{\varphi_1}\ket[b]{\chi_1} +
\lambda_2\ket[ac]{\varphi_2}\ket[b]{\chi_2}$. The Schmidt
coefficients, $\lambda_1$ and $\lambda_2$, determine the non-local
properties of the state with respect to this partition, including
entanglement of the middle qubit $b$ with the rest. Meanwhile, the
entanglement of the reduced state of the end two qubits, $\rho_{ac}$,
can be measured by the concurrence, denoted $C_{ac}$.
Following~\cite{AVBM01}, the state of particles $a$ and $b$ can be
represented by a $4\times2$ matrix $X =
(\lambda_1\ket{\varphi_1},\lambda_2\ket{\varphi_2})$. The concurrence
can be calculated from the singular values $\varsigma_1\ge\varsigma_2$
of $A = X^T\Sigma X$, where $\Sigma = \sigma_y\otimes\sigma_y$:
$C_{ac} = \varsigma_1 - \varsigma_2 = \sqrt{\tr A^\dagger A -
  2\abs{\det A}}$. Taking the time-derivative of this and simplifying
the resulting exact expression (see Appendix~\ref{apdx:3qubit} for
details) leads to the following bound on the entanglement rate:
\begin{equation*}
  \frac{\dd C_{ac}^2}{\dd t} \le 8\Matnorm{H}\lambda_1\lambda_2.
\end{equation*}
The factor $\matnorm{H} = \matnorm{H_{ab}}_1 + \matnorm{H_{bc}}_1$
measures the strengths of the interactions. $\matnorm{H}_1 =
\sum_{ij}\abs{H_{ij}}$ denotes the $l_1$ norm, where the Hamiltonians
are written in the product basis $H =
\sum_{ij}H_{ij}\sigma_i\otimes\sigma_j$, and the Pauli matrices
$\sigma_{1,2,3} = \sigma_{x,y,z}$ are defined in the Schmidt basis
$\{\ket{\chi_1}\ket{\chi_2}\}$. (Local terms
$\sigma_i\otimes\identity$ or $\identity\otimes\sigma_j$ in the
Hamiltonian can not alter the entanglement of the system, so do not
contribute).

In the context of entanglement dynamics, the important part of the
relation is the product of Schmidt coefficients $\lambda_1\lambda_2$,
which is a pure-state bipartite entanglement measure. (In fact, up to
a numerical factor, it is the concurrence.) Thus the differential
equation tells us that the entanglement of the middle qubit limits
entanglement generation between the end qubits: not only must $b$ be
entangled for entanglement to be generated between $a$ and $c$
(precisely what was shown \emph{not} to hold for mixed states
in~\cite{CVDC03}), but the \emph{rate} at which it is generated can be
larger the more entangled $b$ is.

At first sight, the inequality may appear too weak, as it does not
seem to imply that the derivative is zero once qubits $a$ and $c$ are
maximally entangled. However, $C_{ac}$ and $\lambda_1\lambda_2$ are
not independent quantities. When $a$ and $c$ are maximally entangled,
they can not be entangled with anything else, thus
$\lambda_1\lambda_2=0$, and the derivative is zero after all.

A complete quantitative description of entanglement creation in the
three-qubit chain would require an equation describing the evolution
of the Schmidt coefficients. However, $\bipart{b}{ac}$ forms a
bipartite, pure-state system. Entanglement creation in bipartite
systems and the evolution of the Schmidt coefficients has been
investigated in~\cite{DVC+01}.

\subsection{Fidelities and entangled fractions}
The three-qubit result can not directly be extended to higher
dimensional systems. Whilst we can restrict the system to pure states
for the same reasons as in the three-qubit case, the reduced density
matrix of the two end particles can again become mixed during the
evolution. And no closed-form expression is known for the entanglement
of formation of mixed states, other than in the two-qubit case.

Before turning to higher-dimensional systems, it is instructive to
consider more carefully the setting in which we wish to investigate
entanglement dynamics.  Entanglement measures are defined in the LOCC
paradigm: local operations and classical communication (LOCC) can only
decrease the entanglement of a state. This is the natural paradigm
when thinking about entanglement from an information-theorist's point
of view, in which entangled states are shared between different
parties who are free to act locally on their part of the state.

But we are considering entanglement dynamics from a physical
standpoint. In a system of interacting particles, it is not clear what
classical communication means. Any transfer of classical information
between particles would still have to take place via the (quantum)
interactions. It could be argued that it makes more sense in this
context to define entanglement in the local-unitary paradigm: any
change to a state due to local terms in the interaction Hamiltonian
should not change the entanglement.

A physical way of measuring entanglement in this paradigm is to use
the fidelity~\cite{Joz94}, which measures the distance between
states~\footnote{The fidelity is not a metric on density operators,
  but it is closely related to one. See
  e.g.~\cite[\S9.2.2]{Nielsen&Chuang}.}. The \emph{entangled fraction}
of a state $\rho$ is then defined as the maximum fidelity with a
maximally entangled state:
\begin{equation*}
  F(\rho) := \max_{\ket{\phi}\in M.E.} \Bra{\phi}\rho\Ket{\phi},
\end{equation*}
where the maximization is over all maximally entangled states
$\ket{\phi}$ in the bipartite Hilbert space of $\rho$. (For two-qubit
states, it is also called the \emph{singlet} fraction.) It
measures how close a given state is to any maximally entangled state,
and is invariant under local unitary operations, as required.

The entangled fraction is also an experimentally relevant quantity.
When trying to engineer an evolution to produce a particular state (a
highly entangled one, for instance), we want to know how close the
actual state is to the desired one --- precisely what is measured by
the fidelity. For example, in teleportation experiments, it is the
entangled fraction of the entangled pair that determines how close the
teleported state is to the original~\cite{HHH99}.

Therefore, in the remainder of this article, we will consider
evolution of the entangled fraction and related quantities. Though it
is a well-motivated quantity to study in its own right, it can also be
used to give upper and lower bounds on entanglement measures such as
the concurrence~\cite{VV02} (and hence entanglement of formation). In
particular, if a state is separable, its entangled fraction is less
than or equal to~$1/n$ (with $n$ the dimension of the smaller of the
two Hilbert spaces making up the bipartite space). Whereas if (and
only if) the entangled fraction is equal to one, the state must be
maximally entangled.

In the final section, we will use our results to derive bounds on how
long it takes to entangle particles when the system starts in a
separable state. In this context, any quantity that takes different
values for separable and maximally entangled states is equally good in
principle: we can bound the time required to change from one value to
the other. The entangled fraction, for example, must increase from
$1/n$ to $1$.

\subsection{General tripartite chains}
The tripartite chain is a prototype for all indirect (bipartite)
entanglement creation. We can always divide a system into three: two
systems that are being entangled, and everything else lumped into one
mediating system. We can then investigate entanglement flow through
this mediating system.

In a general tripartite chain, consisting of systems $A$, $B$ and $C$
of arbitrary dimension, interacting by nearest-neighbour interactions
$H_{AB}$ and $H_{BC}$, the Schmidt decomposition has the form
$\ket[ABC]{\psi} = \sum_i\lambda_i \ket[AC]{\psi_i}\ket[B]{i}$, where
we sort the Schmidt coefficients $\lambda_i$ in descending order. By
re-expressing the entangled fraction as a maximization over
purifications using Uhlmann's Theorem (see Appendix~\ref{apdx:eflow}),
we can derive an exact expression for the time derivative of the
entangled fraction. Simplifying the exact result to separate out the
entanglement dependence yields a relation analogous to the three-qubit
case (see Appendix~\ref{apdx:tripartite} for details):
\begin{equation*}
  \dot{F}(\rho_{AC}) \le 2\Abs{H}\sqrt{F(\rho_{AC})}\,
  \Bigl(\sum_{ij}\lambda_i\lambda_j-\lambda_1^2\Bigr)
\end{equation*}
Again, the factor $\Abs{H} = \Abs{H_{AB}}_\infty +
\Abs{H_{AB}}_\infty$ measures the interaction strengths, independent
of the system state ($\abs{H}_\infty = \max_{ij}\abs{H_{ij}}$ denotes
the $l_\infty$ norm).

The quantity in brackets is closely related to the entangled fraction
of $\ket[ABC]{\psi}$ in the $\bipart{B}{AC}$ partition:
$F(\proj{\psi}) = \tfrac{1}{n}\sum_{ij}\lambda_i\lambda_j$ (with $n$
the smaller of the dimensions of $B$ and $AC$). Subtracting
$\lambda_1^{2}$ re-scales this entangled fraction so that it is zero
when the state is separable. Therefore, the entanglement of $B$ with
the rest of the system limits the rate at which entanglement can flow
through $B$. As in the three-qubit case, the derivative implicitly
goes to zero when systems $A$ and $C$ become maximally entangled,
since they can not then be entangled with $B$.


\section{Flow along networks \label{sec:along}}
In the previous section, we examined entanglement flow through the
middle particle in a tripartite chain, and noted that the results can
be applied to flow through individual particles in general systems, by
viewing the system as a tripartite chain. However, in a large
multipartite system, this approach means lumping many particles
together into single composite particles, hiding much of the
entanglement dynamics. Can we more fully describe entanglement flow in
networks of interacting particles?

In this section, we derive a set of differential equations describing
the entanglement dynamics, analogous to the rate equations for a
chemical reaction. These show that the rate at which entanglement is
created between two sets of particles depends on the existing
entanglement further back along the network. Intuitively, this can be
interpreted as entanglement flowing through the network.

\subsection{Generalized singlet fraction}\label{sec:genSfrac}
As in the previous section, we must first address the problem of how
to measure entanglement in large systems. Even before that, we must
decide \emph{what} entanglement to measure, since multipartite systems
provide a plethora of possibilities. What questions are we interested
in investigating using our putative equations? Perhaps the most
natural goal, given a system of many interacting particles, is to
entangle a particular pair of them: the end qubits in a chain, for
example. We will take this as our motivation for again considering
entangled fractions of the two particles. We will also need to define
a new fidelity-based quantity to measure bipartite entanglement
embedded in larger systems.

First note that, since any maximally entangled state can be reached by
acting with local unitaries on a particular maximally entangled state,
we can of course maximize over unitaries rather than states in the
definition of the entangled fraction:
\begin{equation*}
  F(\rho) = \max_{U_a,U_b}\Bra{\phi} U_a^\dagger\otimes
    U_b^\dagger\,\rho_{ab}\,U_a\otimes U_b \Ket{\phi}
\end{equation*}
We can equally well think of the unitaries as acting on $\rho$ rather
than on the entangled state $\ket[ab]{\phi}$. This suggests an
alternative interpretation of the singlet fraction: as the maximum
fidelity with a particular maximally entangled state (e.g.\ the
singlet) that can be achieved by acting with local unitaries.

Based on this interpretation, we define the \emph{generalized singlet
  fraction}, a measure of two-qubit entanglement for bipartite systems
of arbitrary dimension (it can be extended in the obvious way to
measure general bipartite entanglement~\cite{note:genSfrac}):
\begin{equation}\label{eq:genSfrac}
  F(\rho_{AB}) = \max_{U_A,U_B} \Bra{\phi}\tr_{\!/\!ab}
  (U_A\otimes U_B\;\rho_{AB}\;U_A^\dagger\otimes U_B^\dagger)
  \Ket{\phi},
\end{equation}
where $a$ and $b$ are qubit systems embedded in $A$ and $B$
respectively, $\ket[ab]{\phi}$ is the singlet state, and the notation
$\tr_{\!/\!ab}$ indicates the partial trace over all systems
\emph{other} than $a$ and $b$. It measures the maximum fidelity with
the singlet achievable by local unitaries.

Note that, in two-qubit systems, this generalized singlet fraction
reduces to the usual singlet fraction. For any system, it takes values
between 0 and 1, and for separable states it is less than or equal to
$1/2$. Also, from the definition, if $A$ and $B$ are subsystems of
$A'$ and $B'$, so that $\rho_{AB} = \tr_{\!/\!AB}(\rho_{A'B'})$, then
$F(\rho_{AB}) \le F(\rho_{A'B'})$.

\subsection{Entanglement rate equations \label{sec:eflow}}
We are now ready to state our main result: a set of coupled
differential equations describing entanglement flow in networks of
interacting particles. For simplicity, we assume that among the set
of interacting particles $S$, there are at least two qubits $a$ and
$b$, the premise being that we are interested in entangling these.
(The results can easily be generalized: see~\cite{note:genSfrac}
and~Conclusions.) Let $A$ and $B$ be disjoint subsets of $S$. The
equations describe the rate at which the generalized singlet fraction
of $\rho_{AB}$ can increase.

\begin{figure}[!htbp]
  \centering
  \includegraphics{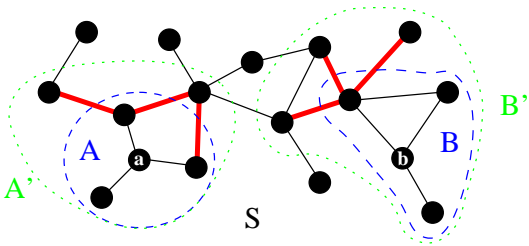}
  \caption{A network of interacting particles, showing interactions
    and sets defined in the entanglement rate equations. Interactions
    `crossing the boundaries' of $A$ or $B$ are indicated by thicker
    lines.}
    \label{fig:network}
\end{figure}

Define $A'$ and $B'$ to be the sets of particles directly connected by
an interaction to at least one particle in $A$ or $B$ respectively
(i.e.\ $A'$ is the set of particles at most `one-away' from $A$, thus
$A\subseteq A'$; see Fig~\ref{fig:network}). If $A'$ and $B'$ are
disjoint (as in Fig.~\ref{fig:network}), then the time derivative of
the generalized singlet fraction is bounded by
\begin{subequations}
\label{eq:eflow}
\begin{equation}\label{eq:eflow1}
  \dot{F}(\rho_{AB}) \le
  2\Matnorm{H}\sqrt{F(\rho_{AB})}\sqrt{F(\rho_{A'B'})-F(\rho_{AB})},
\end{equation}
whilst if $A'$ and $B'$ have one or more particles in common, then
\begin{equation}\label{eq:eflow2}
  \dot{F}(\rho_{AB}) \le
  2\Matnorm{H}\sqrt{F(\rho_{AB})}\sqrt{1-F(\rho_{AB})}.
\end{equation}
\end{subequations}
The factor $\Matnorm{H}$ is a sum of strengths of those interactions
that connect a particle in $A$ or $B$ to one outside $A$ or $B$
respectively (i.e. interactions that `cross the boundary' of $A$ or
$B$; see Fig.~\ref{fig:network}):
\begin{equation*}
  \Matnorm{H} =
  \sum_{\mathclap{\substack{i\in A,j\notin A;\\i\in B,j\notin B}}}
  \Matnorm{H_{ij}}_{HS},
\end{equation*}
where $\Matnorm{\,\bullet}_{HS}$ denotes the Hilbert-Schmidt norm.

The first step in the proof of these \emph{entanglement rate
  equations} to rewrite the generalized singlet
fraction~\eqref{eq:genSfrac} in terms of purifications of $\rho_{AB}$
using Uhlmann's theorem (Appendix~\ref{apdx:eflow}). This leads to the
following exact expression for the derivative of the generalized
singlet fraction:
\begin{equation*}
  \dot{F}(\rho_{AB}) = \sqrt{F(\rho_{AB})}\cdot
  \max_{\substack{U_A,U_B\\\ket{\chi}}}\frac{1}{i}
  \sum_{\mathrlap{\substack{i\notin A(B)\\j\in A(B)}}}
  \Bra{\varphi}H_{ij}\Ket{\psi} - \Bra{\psi}H_{ij}\Ket{\varphi}
\end{equation*}
where $\ket{\psi}$ is a purification of $\rho_{AB}$, $\ket{\chi}$ is
an extension of the singlet state to the Hilbert space of
$\ket{\psi}$, and $\ket{\varphi} = U_A^\dagger\otimes
U_B^\dagger\ket{\chi}$. Using Lemma~\ref{lem:real_imag}
(Appendix~\ref{apdx:eflow}), the terms inside the sum can be bounded
by $\frac{1}{i}(\Bra{\varphi}H_{ij}\Ket{\psi} - \hc) \le
2\Matnorm{H_{ij}}_{\mathrm{HS}} \sqrt{(\tr\Abs{X_{ij}})^2 - (\tr(\real
  X_{ij}))^2}$, where $X_{ij} = \tr_{\!/\!ij}\Ketbra{\psi}{\varphi}$.
Finally, the quantities under the square-root can be related to
generalized singlet fractions: $(\tr(\real X_{ij}))^2 = F(\rho_{AB})$
and $\tr(\Abs{X_{ij}})^2 \le F(\rho_{A'B'})$, which concludes the
proof.  (The proof is given in full detail in
Appendix~\ref{apdx:eflow}).

We can gain some insight into entanglement dynamics by considering the
qualitative meaning of the rate equations, before thinking about
solving them. They divide a network of interacting particles into
pairs of concentric sets, surrounding qubits $a$ and $b$. For example,
in Fig.~\ref{fig:network} there are three such pairs: the qubits $a$
and $b$ themselves, the sets labeled $A$ and $B$, and those labeled
$A'$ and $B'$. The rate equations tell us that entanglement must first
build up between the largest sets, before it can cascade down
successively smaller ones, finally reaching the two qubits (just as in
a chemical reaction, intermediate compounds in the reaction mechanism
must be created before the final product is reached). What is more,
the rate at which the entanglement flows from one level to the next
depends on the difference in entanglement between the two levels
(somewhat like the rate of a reversible chemical reaction, which
depends on the difference between the concentrations of reactants and
products; or like flow in fluids, in which the flow rate depends on
the pressure difference).

The number of pairs of sets is equal to half the `interaction
distance' of the two qubits (rounded down to the nearest integer),
i.e.\ half the smallest number of links in the network needed to
connect $a$ to $b$ (in Fig.~\ref{fig:network}, their interaction
distance is 5). A generalized singlet fraction can be defined on each
pair of sets, along with an accompanying rate equation describing its
evolution. Therefore any network has the same rate equations as a
chain whose length is equal to the interaction distance, and whose
interaction strengths along each link of the chain equal the factors
$\matnorm{H}$; all entanglement flow is equivalent to flow along a
chain. This is qualitatively similar to results from quantum random
walks, in which a quantum walk over a network is equivalent to a
quantum walk along a chain~\cite{FG98}.

Note that the factor $\matnorm{H}$ in the rate equations
indiscriminately includes all interactions that cross the boundary.
We might expect different interactions to contribute differently,
depending on their location in the network. In fact, in
Appendix~\ref{apdx:eflow}, we derive a more general version of the
rate equations, which accounts for each possible interaction pathway
separately, and can therefore take into account the different roles
different particles play in the entanglement dynamics, due to their
differing connectivity. The inequality in the corresponding rate
equations is therefore tighter, but it leads to exponentially (in the
number of particles) more equations describing the entanglement
dynamics of a system, and gives a less intuitive picture of
entanglement flow. We will find that the simpler form given here is
sufficient to derive a number of interesting results.

\subsection{Limits from the rate equations \label{sec:limits}}
It is straightforward to prove inductively that the curves produced by
saturating the inequalities in the rate
equations~(\ref{eq:eflow1},\ref{eq:eflow2}) constitute upper bounds on
the evolution of the generalized singlet fractions; i.e.\ the fastest
possible evolution allowed by the rate equations is that which
saturates the rate equations at each point in time.

If the interaction distance between the qubits we intend to entangle
is $d$, then the full set of rate equations involve $\lfloor d/2
\rfloor$ generalized singlet fractions, which we will denote $F_k(t)$,
$k=1\dots\lfloor d/2 \rfloor$.  ($\lfloor\bullet\rfloor$ denotes
rounding down to the nearest integer.) We number them such that
$F_{\lfloor d/2\rfloor}$ is the singlet fraction of the two qubits.
If we define $F_0=1$, then the evolution of each $F_k(t)$ is described
by equation~\eqref{eq:eflow1}.

Let $f_k(t)$ be the curves that saturate the rate equations, i.e.
$f_k(t)$ is the solution to
\begin{equation*}
  \dot{f_k} = 2\Matnorm{H}\sqrt{f_k}\sqrt{f_{k-1}-f_k},
\end{equation*}
with $f_0=1$. (For simplicity, we can take all coupling strengths
$\matnorm{H}$ to be 1.)  Assume that $f_k(t)$ is an upper bound on
$F_k(t)$, i.e.\ $f_k(t) \ge F_k(t)$ for all $t$. If $f_{k+1}(t)$ is
\emph{not} an upper bound on $F_{k+1}(t)$, then $F_{k+1}(t)$ must
cross it at some point. If this occurs at $t=t_0$, then $F_{k+1}(t_0)
= f_{{k+1}}(t_0)$ and $\dot{F}_{k+1}(t_0) >
\dot{f}_{k+1}(t_0)$~\footnote{It is also possible that at $t_0$, the
  first derivatives are equal: $\dot{F}_{k+1}(t_0) =
  \dot{f}_{k+1}(t_0)$, but a higher order derivative of $F_{k+1}$ is
  larger than that of $f_{k+1}$. In this case, we can find a new point
  $t_0+\epsilon$, infinitesimally close to $t_0$, at which the
  original conditions hold: $F_{k+1}(t_0+\epsilon) =
  f_{k+1}(t_0+\epsilon)$ and $\dot{F}_{k+1}(t_0+\epsilon) >
  \dot{f}_{k+1}(t_0+\epsilon)$. The proof can then be applied at this
  new point.}. But $F_{k+1}(t)$ must still satisfy the inequality in
equation~\eqref{eq:eflow1}. Thus
\begin{align*}
  \dot{F}_{k+1} &\le
  2\sqrt{F_{k+1}(t_0)}\sqrt{F_k(t_0)-F_{k+1}(t_0)}\\
  &\le 2\sqrt{f_{k+1}(t_0)}\sqrt{f_k(t_0)-f_{k+1}(t_0)} =
  \dot{f}_{k+1}(t_0),
\end{align*}
which contradicts the assumption that $F_{k+1}(t)$ crosses
$f_{k+1}(t)$ at $t_0$. Thus if $f_k(t)$ is an upper bound, then so is
$f_{k+1}(t)$.

\begin{figure}[!htbp]
  \centering
  \includegraphics[scale=0.6]{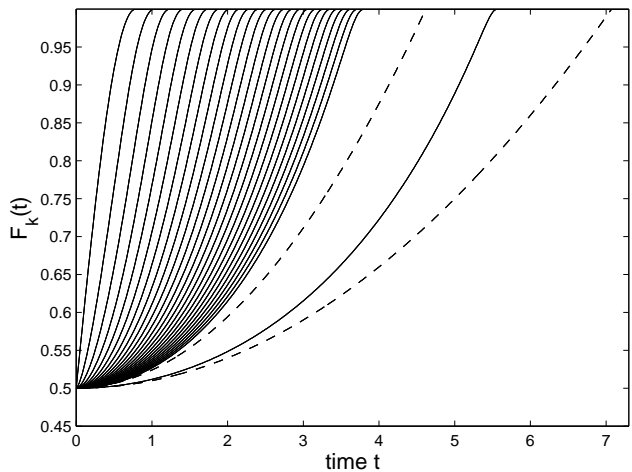}
  \caption{Numerically calculated generalized singlet fraction curves
    $f_k(t)$ saturating the inequalities in the entanglement rate
    equations~(\ref{eq:eflow1},\ref{eq:eflow2}). The final solid curve
    is for $k=50$, i.e.\ the singlet fraction of the end qubits in a
    separated by an interaction distance $d=100$. (The dashed curves
    show the corresponding upper and lower bounds $u_k(t)$ and
    $l_k(t)$ for $k=50$, from subsection~\ref{sec:bounds}).}
  \label{fig:saturate}
\end{figure}

The initial step in the induction (that $f_1(t)$ is an upper bound)
follows from the second of the rate equations~\eqref{eq:eflow2}, and
the fact that the generalized singlet fraction is upper bounded by 1.
Fig.~\ref{fig:saturate} shows numerically calculated curves $f_k(t)$
saturating the rate equations.

\section{How fast can entanglement be created? \label{sec:how_fast}}
How fast can entanglement be generated in a system of interacting
particles? The question is both theoretically interesting, and
experimentally important. Many quantum information processing tasks
require entanglement, and the faster this can be produced, the less
the system will suffer from decoherence. Quantum computing algorithms
often generate large amounts of entanglement during their execution,
so determining how fast entanglement can be generated can also provide
bounds on algorithm complexity~\cite{Amb02}.

In this section, we briefly review some existing entanglement
generation schemes in the context of the entanglement rate equations
derived in the previous section. We then investigate what universal
limits the rate equations put on how fast entanglement can be
generated, or more precisely, how the time required to entangle two
particles scales with the size of the system.

\subsection{Entanglement generation schemes \label{sec:generation}}
How fast entanglement can be generated depends, of course, on how we
are able to manipulate the system. For definiteness, consider
entanglement generation in a qubit chain. It turns out that
measurement is a very powerful resource. If we are able to carry out
local operations on any qubit, including local measurements and
classical communication of the outcomes, then the end qubits in a
chain can be maximally entangled in a time \emph{independent} of the
length of the chain. Though not discussed in the context of
entanglement generation, Briegel and Raussendorf~\cite{BR2001} showed
that a cluster state can be created in a chain in constant time, and
local measurements on a cluster state allow a Bell-state to be
projected out on any desired pair of qubits, including the end
pair~\cite{VPC04}.

The constant scaling assumes we neglect the the time required for
classical communication of the measurement outcomes to the ends of the
chain. This can be justified on theoretical grounds, since classical
communication can not create entanglement, and it makes sense to
consider the interactions as the resource. In many physical
implementations, it is also reasonable on pragmatic grounds: classical
communication is usually much easier to implement than quantum
processes. However, if the interactions are really the \emph{only}
non-local resource, then classical communication must also be
implemented via the chain, and local measurements are of no benefit,
which is equivalent to the local-control scenario described below.
This might be the relevant scenario, for instance, for quantum
computers.

If we can apply local unitary operations on any qubit in the chain,
but \emph{not} measurements, then we can efficiently simulate
evolution under any Hamiltonian (this is true for general systems of
interacting particles, not just for qubit
chains~\cite{NBD+02,WRJB02}). Again, it is reasonable to discount
local resources, which in this scenario means neglecting the time
required to carry out the local unitaries (the `fast local unitary'
approximation). And again, this can also be justified on physical
grounds, since local unitaries are typically much faster than
interactions.

\begin{figure}[!htbp]
  \centering
  \includegraphics[scale=0.6]{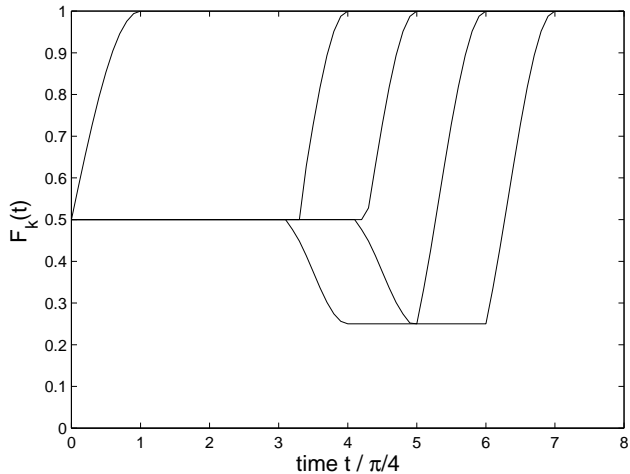}
  \caption{Entanglement dynamics in the entanglement generation
    protocol based on Khaneja and Glaser's state transfer
    scheme~\cite{KG02}, for a chain of 10 qubits. Successive curves
    show the evolution of generalized singlet fractions $F_1$ through
    $F_5$, numbered as in subsection~\ref{sec:limits}.}
  \label{fig:khaneja}
\end{figure}

Khaneja and Glaser have developed an interesting protocol for state
transfer in this scenario~\cite{KG02}, in the context of NMR
spectroscopy, which can easily be transformed into an entanglement
generation protocol. First the middle qubits are entangled, then the
state of each middle qubit is encoded into a three-qubit state. The
encoded states are transferred along the chain towards the ends, where
they are decoded again. The protocol requires local unitaries to be
applied at discrete times. The evolution of the generalized singlet
fractions is shown in Fig~\ref{fig:khaneja}, clearly reflecting the
fact that the protocol is based on moving states step-by-step along
the chain. It achieves a surprising three-fold speedup over the
trivial swapping protocol for entanglement generation in a chain
(entangle the middle qubits; move to the ends by swapping), though the
scaling of the time with the length of the chain is still linear, as
in the trivial protocol. In the next subsection, we will use the
entanglement rate equations to derive a lower bound on the scaling in
this local-control scenario.

\begin{figure}[!htbp]
  \centering
  \includegraphics[scale=0.6]{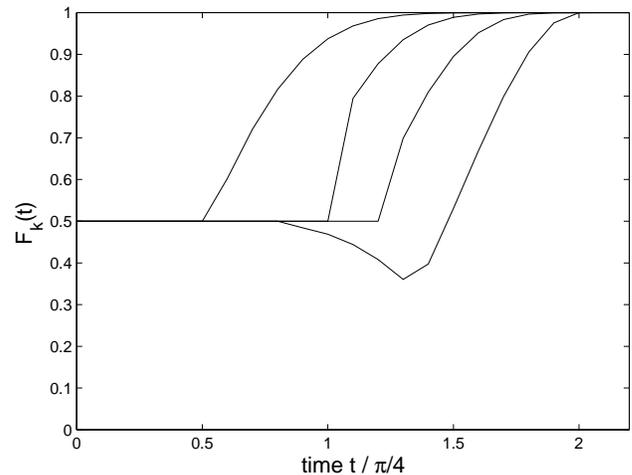}
  \caption{Entanglement dynamics in the entanglement generation scheme
    for odd chain lengths from Ref.~\cite{YLB03}, here for 9 qubits.
    Successive curves show the evolution of generalized singlet
    fractions $F_1$ through $F_4$, numbered as in
    subsection~\ref{sec:limits}. (Note that times are not comparable
    to those in Fig.~\ref{fig:khaneja}, since interaction strengths
    in~\cite{YLB03} are not normalized).}
  \label{fig:leung}
\end{figure}

Finally, we may have no local control over the qubits, only retaining
the ability to switch on interactions in the entire chain, and switch
them off at some later time. Christandl \etal.\ developed a
state-transfer protocol for qubit chains in this
scenario~\cite{CDEL03}, and Yung \etal.\ have given a simple extension
to entanglement generation~\cite{YLB03}. The only local control
required is fixing the coupling strengths between different qubits,
which must be inhomogeneous. Fig.~\ref{fig:leung} shows the
entanglement dynamics for the odd chain-length protocol of
Ref.~\cite{YLB03} --- very different to that of Fig.~\ref{fig:khaneja}.
If the strongest coupling strength is normalized to some fixed value,
then the time to create a maximally entangled pair again scales
linearly with the length of the chain~\cite{note:no_control}.

Osborne and Linden have also developed a protocol for state transfer
in qubit chains, which could be adapted to entanglement generation,
involving limited local control over a vanishingly small (in the limit
of large chain lengths) number of qubits at each end of the
chain~\cite{OL03}.

\subsection{Bounds on entanglement generation \label{sec:bounds}}
In this subsection, we will use the entanglement rate equations to
derive a lower bound on how the time to create a maximally entangled
state scales with the size of the system.

Unfortunately, the set of differential equations defined by the rate
equations~(\ref{eq:eflow1},\ref{eq:eflow2}) has no known closed-form
analytic solution (at least, none that we could find in the
literature). Solving numerically can provide numerical bounds on the
time required for entanglement generation (see
Fig.~\ref{fig:saturate}). The interesting question, though, is how
this time scales with the size of the system (for instance, the length
of a chain), which requires an analytic result.

For simplicity, we will derive a bound on the scaling of the time to
entangle the end qubits in a chain of length $L$. The $\lfloor
L/2\rfloor$ generalized singlet fractions $F_k$ will be numbered such
that $F_{\lfloor L/2 \rfloor}$ is the singlet fraction of the end two
qubits. We assume all interaction strengths are equal to 1, and that
the chain is initially in a completely separable pure state (thus
$F_k(t=0) = 1/2$ for all $k$). The result can easily be generalized to
different interaction strengths, and indeed to general networks of
particles (c.f.\ discussion in subsection~\ref{sec:eflow}).

We are interested in the time at which $F_{\lfloor L/2\rfloor}$ (the
singlet fraction of the two end qubits) reaches 1, as a function of
$L$.  Though the rate equations do not have an analytic solution, we
can inductively prove a bound on the scaling of this time with $L$,
using an argument similar to that used in subsection~\ref{sec:limits}.

There, we showed that the curves obtained when the inequalities in the
rate equations are saturated give upper bounds on the evolution of the
generalized singlet fractions. We can use the same argument to prove
that we still get upper bounds if we weaken the inequalities, using
the fact that $F_k(t)\le 1$, and instead solve
\begin{equation*}
  \dot{f}_k(t) = 2\sqrt{f_{k-1}(t) - f_k(t)}.
\end{equation*}
We can use the argument a third time to prove that if $u_k(t)$ is an
upper bound on the new $f_k(t)$, then the solution $u_{k+1}(t)$ to
\begin{equation}\label{eq:UdiffE}
  \dot{u}_{k+1}(t) = 2\sqrt{u_k(t) - u_{k+1}(t)}
\end{equation}
is an upper bound on $f_{k+1}(t)$. I.e.\ we have $u_k(t) \ge f_k(t)
\ge F_k(t)$. (As concerns boundary conditions, we simply require that
$u_{k+1}(0) \ge f_{k+1}(0) = F_{k+1}(0) = 1/2$.)

Now assume there is a $u_k(t)$ of the form
\begin{equation}\label{eq:Uform}
  u_k(t) = \frac{t^2}{a_k} + \frac{1+\epsilon}{2}
\end{equation}
that is an upper bound on $f_k(t)$ for some positive constants $a_k$
and $\epsilon$. The differential equation for $u_{k+1}(t)$ then has a
solution of the same form as $u_k(t)$ (as can be seen by direct
substitution), with $a_{k+1}$ given by the recursion relation
\begin{equation*}
  a_{k+1} = \frac{a_k}{2} + \frac{a_k}{2}\sqrt{1+\frac{4}{a_k}}.
\end{equation*}
Since $u_{k+1}(0) = (1+\epsilon)/2$, which is greater than the initial
condition $f_{k+1}(0) = F_{k+1}(0) = 1/2$, $u_{k+1}(t)$ is an upper
bound on $f_{k+1}(t)$ by the argument above.

All that remains is the initial step in the induction: that there is
indeed a bound $u_1(t)$ on $f_1(t)$ with the form assumed
in~\eqref{eq:Uform}, for some constants $a_1$ and $\epsilon$.
Fortuitously, the differential equation~\eqref{eq:eflow2} for $F_1(t)$
(the generalized singlet fraction of the entire chain, split into two
halves) \emph{can} be solved analytically when the inequality is
saturated (and without weakening the inequality). The solution has the
form
\begin{equation*}
  f_1(t) = \sin^2(t+\phi)
\end{equation*}
with $\phi$ an arbitrary constant. There is also a trivial solution:
$f_1(t) = 1$. Since the chain starts in a completely separable pure
state, the initial condition is $f_1(0) = 1/2$, and the solution we
require is
\begin{equation*}
  f_1(t) = \begin{cases} \sin^2(t+\pi/4) & t \le \pi/4\\ 1 & t > \pi/4
  \end{cases}
\end{equation*}
The two parts to the solution merely reflect the fact that once the
generalized singlet fraction has reached its maximum value of 1, there
is nothing to be gained by further interaction, and the interactions
affecting $F_1$ (namely the interactions in the middle of the chain)
should be switched off.

Knowing an explicit solution for $f_1(t)$, it is easy to find a bound
$u_1(t)$ with the appropriate form. To make the algebra simpler, we
can upper-bound $f_1(t)$ by $t+1/2$. Thus a $u_1(t)$ with the form
given in~\eqref{eq:Uform} that satisfies $u_1(t) \ge t+1/2$ will
suffice to complete the proof. This leads to the relation $a_1\le
2\epsilon$. Any positive $a_1$ and $\epsilon$ satisfying this will
give an appropriate $u_1(t) \ge f_1(t) \ge F_1(t)$, and will guarantee
that $u_1(0) = (1+\epsilon)/2 \ge f_1(0) = F_1(0) = 1/2$. Therefore we
have shown that an upper bound on $F_1(t)$ with the appropriate form
exists, which completes the proof. For neatness, we can let $\epsilon
\rightarrow 0$, so that $u_k(0) \rightarrow F_k(0) = 1/2$ and
$a_0\rightarrow 0$ (as used to give the curve $u_{50}(t)$ shown in
Fig.~\ref{fig:saturate}).

Solving $u_{\lfloor L/2\rfloor}(t) = 1$ gives a \emph{lower} bound on
the time required for $f_{\lfloor L/2\rfloor}$ to reach 1, which is
itself a lower bound on the time $T_{\mathrm{ent}}$ required for the
singlet fraction of the end two qubits $F_{\lfloor L/2 \rfloor}$ to
reach 1, or equivalently, for the end qubits to become maximally
entangled.

\begin{figure}[!htbp]
  \centering
  \includegraphics[scale=0.6]{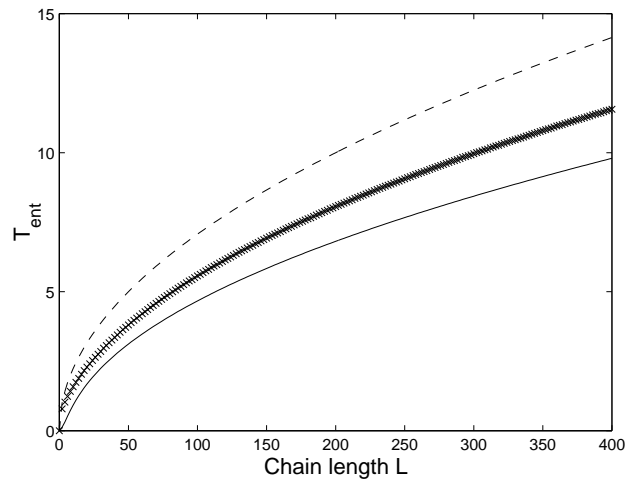}
  \caption{Scaling with chain-length $L$ of the time $T_\mathrm{ent}$
    required to create a maximally entangled state between the ends.
    The points show numerical results obtained by saturating the rate
    equations~(\ref{eq:eflow1},\ref{eq:eflow2}). The solid and dashed
    curves show the analytic lower and upper bounds, $T_{\mathrm{ent}}
    \ge \sqrt{\lfloor L/2 \rfloor/2}$ and $T_{\mathrm{ent}}
    \le \sqrt{\lfloor L/2\rfloor}$ respectively.}
  \label{fig:scaling}
\end{figure}

We are interested in the scaling of $T_{\mathrm{ent}}$ for large chain
lengths, when $a_k$ becomes large. Rather than solving $u_{\lfloor L/2
  \rfloor}(t) = 1$ explicitly to obtain the bound, we can Taylor
expand the square-root in the recursion relation to show that it
asymptomatically approaches $a_k = a_{k-1} + 1$, or equivalently $a_k
= a_1 + k$, as $k \rightarrow \infty$. Thus for large $L$, the bound
tends to
\begin{equation*}
  T_{\mathrm{ent}} \ge \sqrt{\frac{\lfloor L/2 \rfloor}{2}},
\end{equation*}
a square-root scaling with chain length (see Fig.~\ref{fig:scaling}).

We have loosened many an inequality during the proof of the
square-root bound. Could the rate equations give a tighter bound? We
can use essentially the same proof with the inequalities reversed to
prove that a square-root bound is the best that can be obtained.

Instead of using $F_k(t) \le 1$ to weaken the inequality right at the
beginning, we use $F_k(t) \ge 1/2$, which is valid when the $F_k(t)$
saturate the inequalities in the rate equations, i.e.\ when
$F_k(t)=f_k(t)$. Then, solutions of
\begin{equation*}
  \dot{l}_k(t) = \sqrt{2}\sqrt{l_{k-1}(t) - l_k(t)}
\end{equation*}
are \emph{lower} bounds on $f_k(t)$. We can rescale the time $\tau =
t/\sqrt{2}$ to so that the differential equation for $l_k(\tau)$ has
the same form as that for $u_k(t)$ in the previous proof
(Eq.~\ref{eq:UdiffE}). Assuming solutions of the form $l_k(\tau) =
\tau^2/a_k$, solving the resulting recursion relation, and proving
there is a \emph{lower} bound on $f_1(t)$ of the appropriate form,
leads to an \emph{upper} bound on the scaling, for any evolution
saturating the rate equations. For large chain lengths, the bound
tends to $T_{\mathrm{ent}} \le \sqrt{\lfloor L/2\rfloor}$ --- also a
square-root scaling. Therefore, the square-root bound we have derived
is, up to a $\sqrt{2}$ numerical factor, the best that can be obtained
from the entanglement rate equations (see Fig.~\ref{fig:scaling}).

How does our bound compare with the entanglement generation protocols
described in the previous subsection? The generalized singlet
fractions evolve quite differently in those protocols, compared to the
evolution that would saturate the rate equations (compare
Figs.~\ref{fig:khaneja} and~\ref{fig:leung} with
Fig.~\ref{fig:saturate}). All existing protocols that we know of scale
linearly with the length of the chain --- no better than the trivial
swapping protocol (entangle the middle qubits; move to the ends by
swapping). It is an interesting open problem to determine whether any
protocol can achieve a square-root scaling, or whether the bound
derived via the rate equations is too weak and can not be saturated
(which would suggest some improvement on the rate equations might be
possible).


\section{Conclusions}
We have investigated entanglement flow, both through individual
particles and along networks of interacting particles. In both cases,
we have derived differential equations relating the rate of
entanglement generation to the existing entanglement in the
system.

Entanglement flow through a particle is limited by the entanglement of
that particle with the rest of the system, providing the system is in
a pure state. (Previous work~\cite{CVDC03} has already shown that the
entanglement can be transmitted by a particle without that particles
becoming entangled at all, if the system is in a mixed state.)

To describe entanglement flow along general networks of interacting
particles, we have derived a set of entanglement rate equations,
analogous to the rate equations for a chemical reaction. These can
intuitively be interpreted as describing a flow of entanglement along
the network. We have used the rate equations to prove a square-root
lower bound on the scaling with system size of the time required to
create a maximally entangled state, and compared this to existing
entanglement generation protocols. Whether this bound is achievable,
or whether the rate equations can be improved to give a tighter bound,
remains an interesting open problem.


The entanglement rate equations were derived in the context of
two-qubit entanglement creation. However, since they involve
fidelity-based quantities, they can easily be extended to more general
settings. Firstly, the quantities and equations can be extended to
bipartite entanglement generation in arbitrary spaces, by taking
fidelities with a bipartite maximally entangled state in the
appropriate space. Secondly, they can be generalized to the
multipartite setting, by taking fidelities with the desired
multipartite entangled state (e.g.\ a GHZ state), rather than with a
bipartite entangled state.

Together, the results establish a quantitative concept of entanglement
flow in interacting systems. This is of interest as an abstract
concept in itself, but could also be interesting both theoretically
and practically: in the analysis of quantum algorithms, for example,
since these often involve creating large amounts of entanglement
during their operation. Or in physical implementations of quantum
systems, in which it is important to carry out any manipulation
(including entanglement creation) as fast as possible, to beat
decoherence.

\appendix

\section{Three-qubit chain}\label{apdx:3qubit}
To derive the three-qubit result, we use a matrix analysis approach to
calculating the concurrence, developed in~\cite{AVBM01}. Writing the
Schmidt decomposition of the three-qubit system with respect to the
partition $\bipart{b}{ac}$ as $\ket[abc]{\psi} =
\lambda_1\ket[ac]{\varphi_1}\ket[b]{\chi_1} +
\lambda_2\ket[ac]{\varphi_2}\ket[b]{\chi_2}$, we can represent the
state of $ab$ by a $4\times2$ matrix $X =
(\lambda_1\ket{\varphi_1},\lambda_2\ket{\varphi_2})$. The reduced
density matrix is then given by $\rho_{ac} = XX^\dagger$.

The concurrence $C_{ac}$ of $\rho_{ac}$ can be obtained from the
singular values of $A = X^T\Sigma X$, where $\Sigma =
\sigma_y\otimes\sigma_y$~\cite{AVBM01}. In our case, $A$ is a
$2\times2$ matrix (because $\rho_{ac}$ has rank two), with just two
singular values: $\varsigma_1\ge\varsigma_2$. Thus $C_{ac} =
\varsigma_1 - \varsigma_2$. Since $\tr A^\dagger A = \varsigma_1^2 +
\varsigma_2^2$ and $\abs{\det A} = \varsigma_1\varsigma_2$, we can
also write this as
\begin{equation}
  C_{ac}^2 = \tr A^\dagger A - 2\abs{\det A}.
  \label{eq:concurrence}
\end{equation}

To calculate the time-derivative of the concurrence, we must calculate
the derivatives of $\tr(A^\dagger A)$ and $\abs{\det A}$. From its
definition, $\dot{A} = \dot{X}^T\Sigma X + X^T\Sigma\,\dot{X}$.
Meanwhile, $\dd (\tr A^\dagger A)/\dd t = \tr(A^\dagger\dot{A} +
\dot{A}^\dagger A)$, which, after a little algebra, leads to
\begin{equation}\label{eq:trace}
  \frac{\dd(\tr{A^\dagger A)}}{\dd t} = 4\real\left(\tr\bigl(
      \Sigma\rho^*\Sigma\,\dot{X} X^\dagger\bigr)\right).
\end{equation}

Since $A$ is a $2\times2$ matrix, $\det A =
\tr(A\sigma_yA^T\sigma_y)/2$. Thus $\dd(\det A)/\dd t = \tr(\dot{A}
\sigma_y A^T\sigma_y + A\sigma_y\dot{A}^T\sigma_y)/2$ which, after a
little more algebra, gives
\begin{equation}\label{eq:det}
  \frac{\dd (\det A)}{\dd t} = 4\tr\left(X\sigma_y X^T\Sigma
  X\sigma_y \dot{X}^T\Sigma\right).
\end{equation}

The three-qubit chain evolves according to the Hamiltonian $H =
H_{ab}\otimes\identity_c + \identity_a\otimes H_{bc}$. The two-qubit
Hamiltonian $H_{ab}$ has a product decomposition $H_{ab} =
\sum_{ij}a_{ij}\sigma_i\otimes\sigma_j$, where the Pauli matrices
$\sigma_i$ are defined in the $\{\ket{\chi_1},\ket{\chi_2}\}$ basis,
and coefficients $a_{ij}$ are real. Similarly for $H_{bc}$ and
coefficients $c_{ij}$. The Schr\"odinger equation describing the
evolution of the system state $\ket{\psi}$ translates into an equation
for the evolution of $X$:
\begin{equation*}
  \dot{X} = -i\sum_{ij}\left(
    a_{ij}\sigma_i\otimes\identity\,X \sigma_j^T
    + c_{ij}\identity\otimes\sigma_i\,X \sigma_j^T
  \right).
\end{equation*}
We can use this, along with expressions~\eqref{eq:trace}
and~\eqref{eq:det}, in the time-derivative of~\eqref{eq:concurrence}
to obtain an expression for the derivative of the concurrence:
\begin{equation*}
  \frac{\dd C_{ac}^2}{\dd t} = h(H,\Ket{\psi})\lambda_1\lambda_2.
\end{equation*}

The factor $h(H,\ket{\psi}) = \sum_{ij}a_{ij}h_{ij}^a +
c_{ij}h_{ij}^c$ depends on both the interactions and the system state,
and is a rather complicated sum over terms involving $a_{ij}$ and
$c_{ij}$. We define
\begin{align*}
  s_{ij}^k
    &=\bra{\tilde{\varphi_i}}\sigma_k\otimes\identity\ket{\varphi_j}\\
  t_{ij}^k 
    &=\bra{\tilde{\varphi_i}}\identity\otimes\sigma_k\ket{\varphi_j}\\
  o_{ij} &= \braket{\varphi_i}{\tilde{\varphi_j}}\\
  h_{ix}^a
    &= -i(\lambda_1^2 s_{12}^i o_{11} + \lambda_2^2 s_{21}^i o_{22})\\
  h_{iy}^a
    &= \lambda_2^2 s_{21}^i o_{22} - \lambda_1^2 s_{12}^i o_{11}\\
  h_{iz}^a
    &= -i\lambda_1\lambda_2(s_{21}^i o_{12} - s_{12}^i o_{21}),
\end{align*}
and define $h_{ij}^c$ similarly to $h_{ij}^a$, but with the $s_{ij}$'s
replaced by $t_{ij}$'s.  Note that $s_{ii}^k = t_{ii}^k = 0$. (The
tildes denote the spin-flip operation~\cite{Woo98}:
$\ket{\tilde{\varphi}} = \sigma_y\otimes\sigma_y\ket{\varphi^*}$.)
Then
\begin{equation*}
  h(H,\Ket{\psi}) =
  4\real\Bigl(\sum_{ij}a_{ij}h_{ij}^a + c_{ij}h_{ij}^c\Bigr)
  + 4\Bigl\lvert\sum_{ij}a_{ij}h_{ij}^a + c_{ij}h_{ij}^c\Bigr\rvert.
\end{equation*}

However, as we are primarily interested in the dependence on
entanglement (i.e.\ the dependence on the Schmidt coefficients), we
can bound the magnitudes of the $s_{ij}^k$, $t_{ij}^k$ and $o_{ij}$ by
1, and assume all terms sum in phase, giving the bound:
\begin{equation*}
  h(H,\Ket{\psi}) \le \, 8\sum_{ij} \Abs{a_{ij}} + \Abs{c_{ij}},
\end{equation*}
which is independent of the system state, depending only on the
interaction strengths.

\section{Entanglement rate equations}\label{apdx:eflow}
The proof of the entanglement rate equations revolves around Uhlmann's
theorem~\cite{Uhl76,Joz94}, which relates the fidelity of two mixed
states to the fidelity of their purifications:
\begin{thm}[Uhlmann]\label{thm:uhlmann}
  If $\rho$ and $\sigma$ are two states in the same Hilbert space
  $\mathcal{H}$, let $\ket{\psi}$ and $\ket{\varphi}$ be purifications
  of $\rho$ and $\sigma$ into a (in general larger) Hilbert space
  $\mathcal{H\otimes H'}$. Then
  \begin{equation*}
    F(\rho,\sigma) =
    \max_{\ket{\psi},\ket{\varphi}}\Abs{\Braket{\varphi}{\psi}}^2
  \end{equation*}
  where the maximization is over all purifications.
\end{thm}
Since any purification can be transformed into another by a unitary
acting on $\mathcal{H'}$, we can fix one of the purifications and only
maximize over the other one. Also, global phases can be chosen to
ensure the overlap $\braket{\varphi}{\psi}$ is real and positive, so
the absolute value can be dropped.

Recall that the entanglement rate
equations~(\ref{eq:eflow1},\ref{eq:eflow2}) involve two disjoint
subsets, $A$ and $B$, of the entire set of particles $S$, which are
interacting via two-particle interactions $H_{ij}$. We can apply
Uhlmann's theorem to the generalized singlet fraction of $\rho_{AB}$
at time $t$:
\begin{align}
  F_{AB}(t) &= \max_{U_A,U_B} \Bra{\phi}\underbrace{
    \tr_{\!/\!ab}(U_A\otimes U_B\;\rho_{AB}(t)\;
    U_A^\dagger\otimes U_B^\dagger)}_{\sigma_{ab}}\Ket{\phi}\notag\\
  &=\max_{\substack{U_A,U_B\\\ket{\chi}}} \Bra{\chi}U_A\otimes
    U_B\Ket{\psi}^2 \quad \text{by Uhlmann}
    \label{eq:Ft}\\
  &=\Bra{\bar{\chi}}\bar{U}_A\otimes\bar{U}_A\Ket{\psi}^2, \notag
\end{align}
where $\ket{\bar{\chi}}$, $\bar{U}_A$ and $\bar{U}_B$ denote the
particular state and unitaries achieving the maximum. (We are
retaining the unitaries, rather than incorporating them into one of
the purifications, for later convenience. Strictly speaking, they
should be extended to $\mathcal{H}\otimes\mathcal{H'}$ and written
$U_A\otimes U_B\otimes\identity_{\mathrm{rest}}$. In the interests of
economy, we will drop all $\identity_{\mathrm{rest}}$'s.)

The state $U_A\otimes U_B\ket{\psi}$ can be chosen to be any fixed
purification of the two-qubit density operator $\sigma_{ab}$. We use
that freedom to make $\ket{\psi}$ a purification of the overall system
state $\rho_{S}$, which guarantees that $U_A\otimes U_B\ket{\psi}$ is
a purification of $\sigma_{ab}$, as required by Uhlmann's theorem. As
for $\ket{\chi}$, since $\ket{\phi}$ is already pure, it is simply an
extension to $\mathcal{H}\otimes\mathcal{H'}$: $\ket{\chi} =
\ket[ab]{\phi}\ket[\mathrm{rest}]{\vartheta}$ (the maximization then
being over $\ket{\vartheta}$).

If the system evolves under the Hamiltonian $H = \sum_{ij}H_{ij}$ for
an infinitesimal time $\dt$, the state evolves to $\rho_{AB}(t+\dt) =
\tr_{\!/\!AB}(e^{-iH\dt}\rho_S(t)\,e^{iH\dt})$. By writing the density
matrix of the entire system, $\rho_S$, in a product basis for the
partition $\bipart{AB}{\mathrm{rest}}$ and expanding the exponentials
to first order in $\dt$, it is straightforward to show that only
interactions involving at least one particle in $AB$ give a
first-order contribution to the evolution. Therefore, $H$ need only
include that smaller set of interactions. Letting
\begin{equation*}
  \dU = \exp\Bigl(-i\dt\sum_{\mathclap{\substack{i\in S\\j\in AB}}}
  H_{ij}\Bigr)
\end{equation*}
be the resulting (infinitesimal) unitary evolution operator, the
 singlet fraction after the evolution becomes
\begin{equation}\label{eq:Fdt}
  F_{AB}(t+\dt) = \max_{\substack{V_A,V_B\\\ket{\zeta}}}
  \Bra{\zeta}V_A\otimes V_B\cdot\dU\Ket{\xi}^2,
\end{equation}
where we have used Uhlmann's theorem again. The state $\ket{\zeta}$ is
again simply an extension of $\ket{\phi}$ to
$\mathcal{H}\otimes\mathcal{H'}$, and $V_A\otimes V_B\,\dU\ket{\xi}$
can be chosen to be any fixed purification of the two-qubit density
operator
\begin{equation*}
  \tau_{ab} = \tr_{\!/\!ab}\left(V_A\otimes V_B\,\dU\rho_{S}\,
    \dU^\dagger\,V_A^\dagger\otimes V_B^\dagger\right),
\end{equation*}
Again making use of this freedom, and recalling that we chose
$\ket{\psi}$ to be a purification of $\rho_S$, we can choose
$\ket{\xi}$ to be the same state as before: $\ket{\xi} = \ket{\psi}$.

The state $\ket{\bar{\chi}}$ and unitaries $\bar{U}_A$ and $\bar{U}_B$
were defined to be those maximizing expression~\eqref{eq:Ft}. Thus by
definition,
\begin{equation*}
  \Bra{\bar{\chi}} \bar{U}_A\otimes\bar{U}_B\Ket{\psi}
  \ge \Bra{\chi} U_A\otimes U_B\Ket{\psi}
\end{equation*}
for all $\ket{\chi}$, $U_A$ and $U_B$. In particular, this is true for
infinitesimal changes, e.g.\ $\ket{\bar{\chi}}+\dt\ket{\chi^{\bot}}$
where $\ket{\chi^{\bot}}$ is orthogonal to $\ket{\chi}$. Thus
$\bra{\chi^\bot}\bar{U}_A\otimes\bar{U}_B\ket{\psi} \le 0$. However,
if this were strictly negative for some $\ket{\chi^\bot}$, then
$-\ket{\chi^\bot}$ would make it positive. Therefore
$\bra{\chi^\bot}\bar{U}_A\otimes\bar{U}_B\ket{\psi} = 0$. Similarly,
considering infinitesimal changes to the unitaries, we can show that:
\begin{gather}\label{eq:relations}
  \begin{gathered}
    \Bra{\bar{\chi}}\bar{U}_A H_A\otimes \bar{U}_B\Ket{\psi} = 0\quad
    \Bra{\bar{\chi}}\bar{U}_A\otimes \bar{U}_B H_B\Ket{\psi} = 0\\
    \bra{\chi^\bot}\bar{U}_A\otimes\bar{U}_B\ket{\psi} = 0.
  \end{gathered}
\end{gather}

Expression~\eqref{eq:Fdt} for the generalized singlet fraction at time
$t+\dt$ must tend to expression~\eqref{eq:Ft} (the corresponding
expression for time $t$) as $\dt\rightarrow 0$, so $\ket{\zeta} =
\ket{\bar{\chi}}+\dt\ket{\chi^\bot}$ and $V_{A(B)} =
\bar{U}_{A(B)}(\identity + i\dt H_{A(B)})$, where $H_{A(B)}$ is a
Hermitian operator on $A(B)$. Using this, expanding $\dU = \identity -
i\dt H + \order{\dt^2}$ (where $H$ is the sum of interactions
involving at least one particle in $A$ or $B$), and making use of
relations~\eqref{eq:relations}, we have
\begin{equation*}
  F_{AB}(t+\dt) = \Bra{\bar{\chi}}\bar{U}_A\otimes\bar{U}_B
  (\identity-i\dt H)\Ket{\psi}^2 + \Order{\dt^2}.
\end{equation*}
I.e.\ the state and unitaries maximizing expression~\eqref{eq:Ft} also
maximize~\eqref{eq:Fdt}, to first order in $\dt$.

Hamiltonian $H$ currently includes all interactions involving at least
one particle in $A$ or $B$. By expanding $H$ in the previous
expression as a sum over these two-particle interactions, we can use
the same relations~\eqref{eq:relations} to show that only interactions
crossing the \emph{boundary} of $A$ or $B$ need to be included to give
the generalized singlet fraction to first order in $\dt$:
\begin{equation*}
  F_{AB}(t+\dt) = \Bra{\bar{\chi}}\bar{U}_A\otimes\bar{U}_B\Bigl(
  \identity - i\dt\sum_{\mathclap{\substack{i\notin A(B)\\j\in A(B)}}}
  H_{ij}\Bigr)\Ket{\psi}^2 + \Order{\dt^2}.
\end{equation*}

Now, global phases were chosen to make it real and positive, so
$\bra{\bar{\chi}}\bar{U}_A\otimes\bar{U}_B\ket{\psi} =
\sqrt{F_{AB}(t)}$. Thus, expanding the square in the previous
expression and only retaining first order terms in $\dt$, we arrive at
a first expression for the time-derivative of the generalized singlet
fraction:
\begin{equation}\label{eq:Fdot}
  \dot{F}_{AB}(t) = \sqrt{F_{AB}(t)}\cdot\frac{1}{i}
  \sum_{\mathrlap{\substack{i\notin A(B)\\j\in A(B)}}}
  \Bra{\varphi}H_{ij}\Ket{\psi} - \Bra{\psi}H_{ij}\Ket{\varphi}
\end{equation}
where $\ket{\varphi} =
\bar{U}_A^\dagger\otimes\bar{U}_B^\dagger\ket{\bar{\chi}}$.

To proceed, we will need the following Proposition~\cite{Bhatia} which
we use to prove the subsequent Lemma:
\begin{prop}[Fan-Hoffman] \label{prop:fan-hoffman}
  For any operator $X$, the ordered singular values
  $\sigma^\downarrow_i$ of $X$ are individually greater than or equal
  to the ordered eigenvalues $r^\downarrow_i$ of $\real X =
  (X+X^\dagger)/2$. I.e.\ $\sigma^\downarrow_i \ge r^\downarrow_i$
  $\forall i$.
\end{prop}
Note that the eigenvalues of $\real X$ can be negative, in which case
the \emph{absolute values} of the eigenvalues need not obey the
Proposition.

\begin{lem}\label{lem:real_imag}
  For any operator $X$, $(\tr\abs{X})^2 - (\tr(\real X))^2 \ge
  \tr((\imag X)^2)$, where $\abs{X} = \sqrt{XX^\dagger}$, $\real{X} =
  (X+X^\dagger)/2$, and $\imag{X} = (X-X^\dagger)/2i$.
\end{lem}
\begin{prf}
  Assume initially that $\tr(\real X)$ is non-negative. Defining $P$
  ($N$) to be the set of positive (negative) eigenvalues of $\real X$,
  \begin{align*}
    \sum_{i\ne j}&\left(\sigma_i\sigma_j - r_i r_j\right) \\
      =& \sum_{i\ne j}\sigma_i\sigma_j
        - \sum_{\substack{i,j\in P\\i\ne j}}r_i r_j 
        + \sum_{\substack{i\in P\\j\in N}}\Abs{r_i}\Abs{r_j} \\
      &+ \sum_{\substack{i\in N\\j\in P}}\Abs{r_i}\Abs{r_j}
        - \sum_{\substack{i,j\in N\\i\ne j}}\Abs{r_i}\Abs{r_j} \\
      =& \sum_{\mathclap{\substack{i\,\text{or}\,j\notin P\\i\ne j}}}
        \sigma_i\sigma_j
        + \sum_{\substack{i\in P\\j\in N}}\Abs{r_i}\Abs{r_j}
        + \sum_{\substack{i,j\in P\\i\ne j}}\left(
          \sigma^\downarrow_i\sigma^\downarrow_j
          - r^\downarrow_i r^\downarrow_j
        \right)  \\
      &+\sum_{i\in N}\Abs{r_i}\biggl(
        \sum_{j\in P}\Abs{r_j}
        - \sum_{\substack{j\in N\\j\ne i}}\Abs{r_j}
        \biggr).
  \end{align*}
  The first two terms are clearly positive, the third is positive by
  Proposition~\ref{prop:fan-hoffman}, and the last by the assumption
  that $\tr(\real X) \ge 0$. Thus $\sum_{i\ne j}\left(\sigma_i\sigma_j
  - r_i r_j\right) \ge 0$. Now,
  \begin{align*}
    \bigl(\tr&\Abs{X}\bigr)^2 - \bigl(\tr(\real X)\bigr)^2
      =\Bigl(\sum_i\sigma_i\Bigr)^2 - \Bigl(\sum_i r_i\Bigr)^2\\
    &=\sum_i\sigma_i^2 - \sum_i r_i^2
      + 2\sum_{i\neq j}\left(\sigma_i\sigma_j - r_i r_j\right)\\
    &\ge \sum_i\sigma_i^2 - \sum_i r_i^2\quad \\
    &=\tr\left(XX^\dagger\right) - \tr((\real X)^2) = \tr((\imag X)^2)
  \end{align*}
  where in the last line we have expanded $X = \real X + i\imag X$,
  and used the fact that $\real X$ and $\imag X$ are both Hermitian
  and that the trace of their commutator is 0.
  
  For completeness, we can remove the assumption $\tr(\real X) \ge 0$
  by noting that, if there existed an operator $X$ with $\tr(\real X)
  < 0$ such that the Lemma did not hold, then the operator $-X$ would
  also violate the Lemma. But then $\tr(\real(-X)) \ge 0$, so the
  Lemma must hold for all operators.
\end{prf}

Recall that $H_{ij}$ really means
$H_{ij}\otimes\identity_{\mathrm{rest}}$. Thus
\begin{align}
  \frac{1}{i}\bigl(&\Bra{\varphi}H_{ij}\Ket{\psi} - \hc \bigr)
  = \tr\Bigl(H_{ij}\cdot\frac{1}{i} \left(\tr_{\!/\!ij}
      \Ketbra{\psi}{\varphi} - \hc\right)\Bigr)\notag\\
  &= 2\tr\left(H_{ij}\imag X_{ij}\right)\quad \text{where }
    X_{ij} = \tr_{\!/\!ij}\Ketbra{\psi}{\varphi}\notag\\
  &\le 2\sqrt{\tr{H_{ij}^2}}\sqrt{\tr((\imag X_{ij})^2)}\quad
    \text{by Cauchy-Schwartz}\notag\\
  &\le 2\Matnorm{H_{ij}}_{\mathrm{HS}}
    \sqrt{(\tr\Abs{X_{ij}})^2 - (\tr(\real X_{ij}))^2}
    \label{eq:bracket}
\end{align}
using Lemma~\ref{lem:real_imag} in the last line.
($\matnorm{\bullet}_{\mathrm{HS}}$ denotes the Hilbert-Schmidt norm.)

Finally, we need to relate the quantities under the square-root to
generalized singlet fractions. Firstly, $(\tr(\real X_{ij}))^2 =
(\real (\tr\ketbra{\psi}{\varphi}))^2 = \braket{\varphi}{\psi}^2 =
F_{AB}(t)$, since global phases were chosen to make
$\braket{\varphi}{\psi}$ real and positive.

Secondly, $H_{ij}$ acts on one particle $j$ within $A$ or $B$, and a
particle $i$ outside. If $j$ is in $A$, define the sets $A'_i=A\cup i$
and $B'_i=B$. If it is in $B$, define $A'_i=A$, $B'_i=B\cup i$. We
apply Uhlmann's theorem to the definition of the generalized singlet
fraction for $\rho_{A'_iB'_i}$, and again choose the same state
$\ket{\psi}$ for one of the purifications. So long as $A'_i$ and
$B'_i$ are disjoint, we have
\begin{align*}
  F&_{A'_iB'_i}(t) = \max_{\substack{V_{A'},V_{B'}\\\ket{\zeta}}}
    \Bra{\zeta} V_{A'}\otimes V_{B'} \Ket{\psi}^2\\
  &\ge \max_{V_{A'},V_{B'}} \Bra{\bar{\chi}}
    \bar{U}_A V_{A'}\otimes \bar{U}_B V_{B'} \Ket{\psi}^2\\
  &= \max_{V_{A'},V_{B'}} \Bra{\varphi}V_{A'}\otimes V_{B'}
    \Ket{\psi}^2
    \ge\max_{U_{ij}} \Bra{\varphi}U_{ij}\Ket{\psi}^2\\
  &= \max_{U_{ij}}(\tr(U_{ij}X_{ij}))^2 = \tr(\Abs{X_{ij}})^2,
\end{align*}
where an inequality appears each time we restrict the maximization.
The last line follows from the fact that, for any operator,
$\tr\abs{X} = \max_U\abs{\tr(UX)}$, which is easily proved via the
polar decomposition of $X$. If $A'_i$ and $B'_i$ have a particle in
common (it must be particle $i$ if they do), then the second of the
two inequalities is not valid. We can instead bound
$\tr(\abs{X_{ij}})^2 \le 1$.

Thus, using $(\tr(\real X_{ij}))^2 = F_{AB}(t)$ and
$\tr(\Abs{X_{ij}})^2 \le F_{A'_iB'_i}(t)$ ($A'_i$ and $B'_i$ disjoint)
or $\tr(\Abs{X_{ij}})^2 \le 1$ ($A'_i$ and $B'_i$ overlapping)
in~\eqref{eq:bracket}, and substituting the result in~\eqref{eq:Fdot},
we arrive at a version of the entanglement rate equation:
\begin{equation*}
  \dot{F}_{AB}(t) \le
  2 \sum_{\mathrlap{\substack{i\notin A(B)\\j\in A(B)}}}
  \Matnorm{H_{ij}}_{\mathrm{HS}}\sqrt{F_{AB}(t)\vphantom{F_{A'_i}}}
  \sqrt{F_{A'_iB'_i}(t) - F_{AB}(t)},
\end{equation*}
where $F_{A'_iB'_i}$ is defined to be $1$ if $A'_i$ and $B'_i$
overlap.

To describe entanglement flow in a network of interacting particles,
one such rate equation must be written down for \emph{all} meaningful
generalized singlet fractions that can be defined on the network
(`meaningful' implying that the sets $A$ and $B$ include particles $a$
and $b$ respectively, and are each be made up of `one piece').

Recall that, since $A'_i$ and $B'_i$ are subsets of $A'$ and $B'$ (see
Fig.~\ref{fig:network}), $F_{A'_iB'_i} \le F_{A'B'}$. We can use this
to arrive at the simpler version of the entanglement rate equations
presented in the main text:
\begin{equation*}
  \dot{F}_{AB}(t) \le
  2 \sum_{\mathclap{\substack{i\notin A(B)\\j\in A(B)}}}
    \Matnorm{H_{ij}}_{\mathrm{HS}}
  \sqrt{F_{AB}(t)}\sqrt{F_{A'B'}(t) - F_{AB}(t)}.
\end{equation*}

\section{General tripartite chains}\label{apdx:tripartite}
The first half of the derivation of the entanglement rate equations
given in Appendix~\ref{apdx:eflow} can be re-used in the proof of the
tripartite chain result. Recall that the three systems $A$, $B$ and
$C$ making up the chain are in an overall pure state
$\ket[ABC]{\psi}$, and interact by nearest-neighbour interactions: $H
= H_{AB} + H_{BC}$. As noted in subsection~\ref{sec:genSfrac}, the
 entangled fraction $F_{AC}$ of $\rho_{AC}$ can be expressed as
a maximization over unitaries $U_A$ and $U_C$ rather than states.
Applying Uhlmann's relation, it can be rewritten
\begin{equation*}
  F_{AC} = \max_{\substack{U_A,U_C\\\ket{\chi}}}
  \Bra{\chi}U_A\otimes U_C\Ket{\psi} = \Braket{\varphi}{\psi}.
\end{equation*}
We can choose $\ket{\psi}$ to be the overall system state
$\ket[ABC]{\psi}$, since this is a purification of $\rho_{AC}$.
$\ket{\chi}$ is then an extension of a maximally entangled state
$\ket[AC]{\phi}$ on $\mathcal{H}_{AC}$ to the space
$\mathcal{H}_{ABC}$: $\ket{\chi} = \ket[AC]{\phi}\ket[B]{\zeta}$. We
define $\ket{\varphi} =
\bar{U}_A^\dagger\otimes\bar{U}_C^\dagger\ket{\bar{\chi}} =
\ket[AC]{\bar{\phi}}\ket{\bar{\zeta}}$, where bars denote the
particular unitaries and states achieving the maximum.

Although the rate equations involve the \emph{generalized} singlet
fraction, up to expression~\eqref{eq:Fdot} the derivation in
Appendix~\ref{apdx:eflow} applies equally well to the entangled
fraction. Expression~\eqref{eq:Fdot} then becomes
\begin{equation}\label{eq:FACdot}
  \dot{F}_{AC}(t) = \sqrt{F_{AC}(t)}\cdot\frac{1}{i}
  \left(\Bra{\varphi}H\Ket{\psi} - \Bra{\psi}H\Ket{\varphi}\right),
\end{equation}

Now, writing the state in its Schmidt decomposition for the partition
$\bipart{B}{AC}$, $\ket[ABC]{\psi} = \sum_i\lambda_i
\ket[AC]{\psi_i}\ket[B]{i}$, where we sort the Schmidt coefficients in
descending order: $\lambda_1\ge\lambda_2\ge\dots\ge\lambda_n$.
Extending $\{\ket[B]{i}\}$ to form a complete basis for
$\mathcal{H}_B$, $\ket{\varphi}$ can be written in the product
decomposition $\ket{\varphi} = \ket[AC]{\bar{\phi}}\sum_i
\alpha_i\ket[B]{i}$ (the $\alpha_i$ are complex in general).

We know that $\ket{\varphi}$ maximizes $\braket{\varphi}{\psi} =
\sum_i\alpha_i^*\lambda_i\Braket{\bar{\phi}}{\psi_i}$. Clearly, the
phases of $\alpha_i$ must be chosen to cancel the phases of
$\braket{\bar{\phi}}{\psi_i}$. The relationship between the magnitudes
of the $\alpha_i$'s and $\braket{\bar{\phi}}{\psi_i}$'s can be found
using Lagrange multipliers, with the normalization constraint
$\sum_i\abs{\alpha_i}^2 = 1$, yielding
\begin{equation}\label{eq:alpha}
  \alpha_i = \frac{\lambda_i\Braket{\psi_i}{\bar{\phi}}}
    {\sqrt{\sum_k\lambda_k^2\Abs{\Braket{\bar{\phi}}{\psi_k}}^2}}
\end{equation}

We also know that, for any Hamiltonian acting only on $A$ or $C$,
$\bra{\varphi}H_A\ket{\psi} = \bra{\varphi}H_C\ket{\psi} = 0$
(applying the same reasoning as used to prove
relations~\eqref{eq:relations} in Appendix~\ref{apdx:eflow}). Thus
\begin{equation}\label{eq:i=1}
  \alpha_1^*\lambda_1\Bra{\bar{\phi}}H_{A(C)}\Ket{\psi_i}
  = -\sum_{i\ne1}\alpha_i^*\lambda_i
  \Bra{\bar{\phi}}H_{A(C)}\Ket{\psi_i}.
\end{equation}

Now, the system Hamiltonian $H = H_{AB} + H_{AC}$.
$\Bra{\varphi}H_{AB}\Ket{\psi} = \sum_{ij}\alpha_i^*\lambda_j
\bra{\bar{\phi}}\bra[B]{i}H_{AB}\ket[B]{j}\ket{\psi_j}$. For the
$i=j=1$ terms in the sum, $\bra[B]{1}H_{AB}\ket[B]{1}$ is just some
Hamiltonian acting only on $A$. Similarly for
$\bra[B]{1}H_{BC}\ket[B]{1}$. Thus using~\eqref{eq:alpha}
and~\eqref{eq:i=1},
\begin{align*}
  \Bra{\varphi}H\Ket{\psi} = &\sum_{i\ne1}\alpha_i^*\lambda_i
    \bigl(\Bra{\bar{\phi},i}H\Ket{\psi_i,i}
    - \Bra{\bar{\phi},1}H\Ket{\psi_i,1}\bigr)\\
  &+ \sum_{i\ne j}\alpha_i^*\lambda_i
    \Bra{\bar{\phi},i}H\Ket{\psi_j,j}\\
  = &\sum_{i\ne 1}\lambda_i^2 h_{ii}
    + \sum_{i\ne j}\lambda_i\lambda_j h_{ij},
\end{align*}
where
\begin{align*}
  h_{ii} &= \frac{\Braket{\psi_i}{\bar{\phi}}\bigl(
    \Bra{\bar{\phi},i}H\Ket{\psi_i,i}
    - \Bra{\bar{\phi},1}H\Ket{\psi_i,1} \bigr)}
    {\sqrt{\sum_k\lambda_k^2\Abs{\Braket{\bar{\phi}}{\psi_k}}^2}}\\
  h_{ij} &=
  \frac{\Braket{\psi_i}{\bar{\phi}}\Bra{\bar{\phi},i}H\Ket{\psi_i,j}}
  {\sqrt{\sum_k\lambda_k^2\Abs{\Braket{\bar{\phi}}{\psi_k}}^2}}
  \quad\text{for } i\ne j.
\end{align*}

Using this in~\eqref{eq:FACdot}, and bounding $(h_{ij}-h_{ij}^*)/i \le
2\abs{H} = 2(\abs{H_{AB}}_\infty + \abs{H_{BC}}_\infty)$ (where
$\abs{M}_\infty = \max_{ij}\abs{M_{ij}}$ denotes the $l_\infty$ norm),
we arrive at the final result:
\begin{equation*}
  \dot{F}_{AC}(t) \le 2\Abs{H}\sqrt{F_{AC}(t)}\,
  \Bigl(\sum_{ij}\lambda_i\lambda_j-\lambda_1^2\Bigr).
\end{equation*}

\bibliography{E-flow}

\begin{thebibliography}{33}
\bibitem[{not({\natexlab{a}})}]{note:genSfrac}
\bibinfo{note}{According to the definition given, the generalized singlet
  fraction is only defined for spaces in which qubits can be embedded. It can
  be extended to general spaces if, instead of embedding a qubit, we embed a
  qudit of appropriate dimension. It can also be extended to measure
  multi-partite entanglement by replacing the bipartite maximally entangled
  state with the desired multipartite entangled state (e.g.\ a GHZ state).}
\bibitem[{not({\natexlab{b}})}]{note:no_control}
\bibinfo{note}{Ref.~\cite{CDEL03} states that the time required to entangle the
  end qubits is independent of the length of the chain. For this to be the
  case, the interaction strengths must be increased for longer chains. If the
  interaction strengths are normalized, the scaling is linear.}
\end{thebibliography}

\end{document}